\documentclass[aip,jcp,reprint,amsmath]{revtex4-1}
\usepackage{graphicx}
\usepackage{epstopdf}
\usepackage{dcolumn}
\usepackage{bm}

\begin{document}
	
\author{Hualin Zhan}
\email{hualin.zhan@gmail.com}
\affiliation
{School of Physics, The University of Melbourne, Parkville, VIC 3010, Australia}
\author{Jiri Cervenka}
\affiliation
{Institute of Physics ASCR, v.v.i., Cukrovarnick\'a 10/122, 16200 Praha 6, Czech Republic}
\author{Steven Prawer}
\affiliation
{School of Physics, The University of Melbourne, Parkville, VIC 3010, Australia}
\author{David J. Garrett}
\affiliation
{School of Physics, The University of Melbourne, Parkville, VIC 3010, Australia}

\title
  {Molecular detection by liquid gated Hall effect measurement of graphene}



\begin{abstract}
 The Hall resistance obtained in liquid gated Hall effect measurement of graphene demonstrates a higher sensitivity than the sheet resistance and the gate-source current for L-histidine of different concentrations in the pM range. This indicates that the extra information offered by the liquid gated Hall measurement of graphene can improve the sensitivity of the transistor-based potentiometric biosensors, and it could also be a supplementary method to the amperometric techniques for electrochemically inactive molecules. Further analysis of the system suggests that the asymmetry of the electron-hole mobility induced by the ions in the liquid serves as the sensing mechanism. The calculation on the capacitance values shows that the quantum capacitance is only dominant near the ``Dirac'' point in our system. This conclusion is useful for many applications involving graphene-electrolyte systems, such as bio-sensing, energy storage, neural stimulation, and so on.
\end{abstract}


\maketitle

\section{Introduction}

\begin{figure*}
	\includegraphics[width=\textwidth]{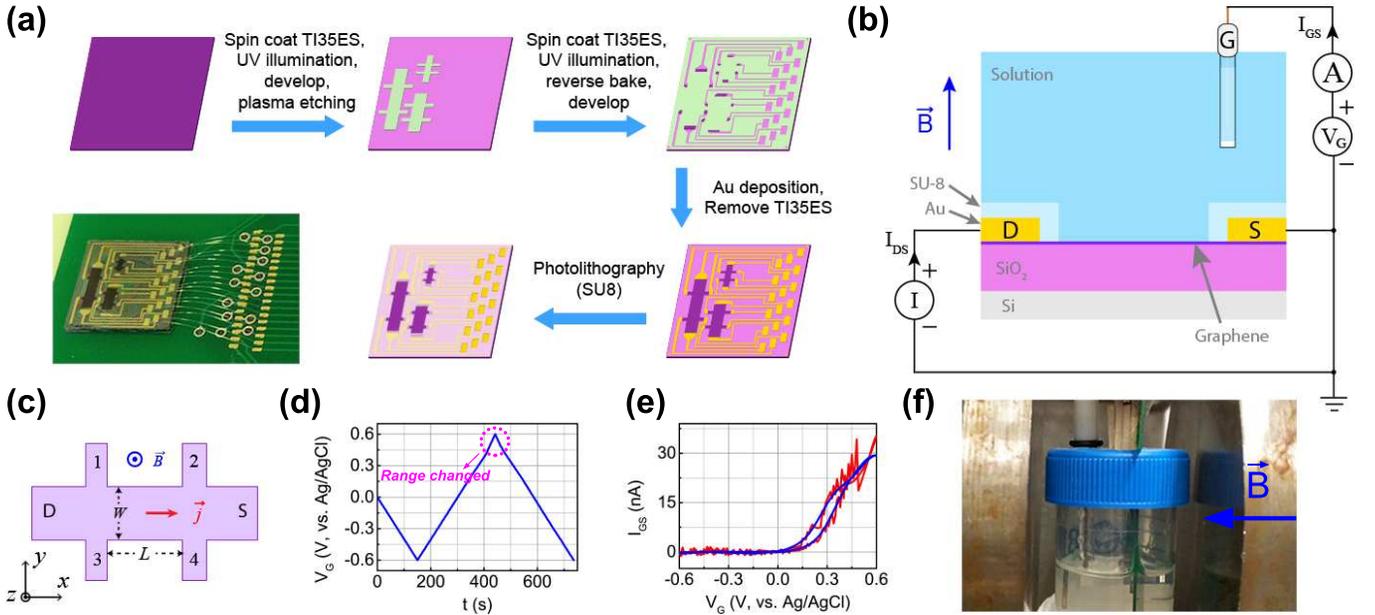}
	\caption{ \textbf{(a)}, Fabrication process of graphene Hall structures and a photo of a device. \textbf{(b)}, Schematic diagrams of the setup of liquid gated Hall measurement, where G is the gate (Ag/AgCl electrode), I, $V_G$ and A are constant current source, gate voltage source, and ammeter, respectively. $I_{GS}$ and $I_{DS}$ are the gate-source current and the drain-source current, respectively. \textbf{(c)}, Illustration of a graphene Hall bar on $SiO_2/Si$ substrate, where D and S represent drain and source, respectively. \textbf{(d)}, $V_G$ as a function of time for the equivalent CV. \textbf{(e)}, $I_{GS}$ as a function of $V_G$ obtained in the CV, where the red and blue lines are the raw and the smoothed data (by Fourier transform), respectively. \textbf{(f)}, A photo of the liquid gated Hall measurement setup.
		\label{fig1}}
\end{figure*}

Detection of molecules in liquid at low concentration has been of great interest for life science applications including medicine, ecology and biology. Developing sensitive and fast biosensors is one of the main goals in these fields\cite{G_bio_science,G_bio_nature}. Current electrochemical bio-sensing techniques are mainly categorized as amperometric or potentiometric, such as Cyclic Voltammetry (CV)\cite{echem_C,G_bio_cv1} or ion-sensitive field effect transistor (ISFET)\cite{G_bio_fet2,G_bio_fet1,G_bio_fet3,NL_Fu,AC_ISFET,AHM_Yan}, respectively. 

CV identifies molecules by measuring the potential at which a species is electrochemically oxidised or reduced, while ISFET measures the electrical resistance change of the channel after interaction with the analyte molecules. CV can only detect molecules which are electrochemically active, i.e. a redox reaction must occur during the measurement. For the molecules which are electrochemically inactive however, it is difficult to identify them by amperometric method. When the resistance change of pristine ISFETs in response to some low concentrated solutions is not observable\cite{G_bio_fet1,NL_Fu}, some pre-treatment of the device to increase the sensitivity is often required, such as introducing defects\cite{G_bio_fet1}, depositing molecular binding sites\cite{AHM_Yan,AFM_lt,Ns_lt}, or growing an extra atomic layer on graphene\cite{NL_Fu}. 

Electrical conductance ($\sigma$) is the product of carrier charge density ($en$) and mobility ($\mu$), $\sigma = en\mu$, where $e$ is the electronic charge and $n$ is the carrier number density. Hence although $\sigma$ does not vary with the analyte concentration in some cases, $n$ and/or $\mu$ could still change evidently. For example, if $n$ increases by a factor of $\alpha$ ($n_2=\alpha n_1$) while $\mu$ decreases by a factor of $\frac{1}{\alpha}$ ($\mu_2=\frac{\mu_1}{\alpha}$), the electrical conductance remains the same ($\sigma_2=\sigma_1$). An ISFET in such a case is not able to detect the molecules as the electrical conductance of the channel does not change. However, the changes of the carrier density and the mobility of the channel induced by the molecules can be captured by the simultaneous measurement of Hall resistance ($R_H$) and sheet resistance ($\rho_{xx}$) during Hall effect measurements. 

This paper presents a very sensitive molecular detection method using Hall effect measurements on graphene devices without pre-treatment (such as functionalization, etc.) where the gate electrode is immersed in the solution containing the analyte of interest. This liquid gated Hall measurement (LGHM) technique allows to exclude contact resistances and directly measure the Hall resistance ($R_H$) and the sheet resistance ($\rho_{xx}$) of the sensor channel (graphene) during the exposure of the sensor to analytes in liquids. The obtained $R_H$ in various gate potentials exhibits a very high sensitivity for a low concentration solution in the pM range, while the conventional methods including CV and ISFET do not show an observable response under the same conditions. Hall measurements of graphene in liquid have been performed for different purposes, such as determining the carrier density of graphene\cite{PNAS_lh}, studying the charge scattering mechanism\cite{NC_lh}, and so on. However, to the best of our knowledge there has been no work published so far on applying this method in bio-sensing in liquid.

Detection of L-histidine and urea are tested in this study because these biomolecules are well known and exist in the organism in low and high concentrations, respectively. L-histidine, as an essential positively charged amino acid (with a dipole moment of 3.6 D) used to synthesize proteins in humans, is of interest to many different fields of applications\cite{app1,app2,app_med,aa_meta}. Clinically, L-Histidine is correlated with the liver metabolism of melancholic patients\cite{app_med,aa_meta} which can appear as an increased histidine concentration in the urine of such patients. It is therefore of interest to detect this molecule and urea. The sensitivity of l-histidine detection is usually in the nM/$\mu \text{M}$ range without modifying the electrodes\cite{sensing1,sensing2} (it should be noted that further improvements can be made by functionalizing the electrodes\cite{sensing3}). Urea is an neutral organic compound (with a dipole moment of 4.56 D) synthesized in many organisms, including livers, of mammals. It plays a key role in carrying waste nitrogen and regulation of water concentration and blood pH in the body. The typical concentration of urea that occur in human blood are in the mM range\cite{physio,urea1,urea2}. The detections of L-histidine and urea are tested in the pM and mM range concentrations, respectively. The sensing mechanism is first experimentally explored by measuring the electronic properties of graphene immersed in DI water and in phosphate-buffered saline (PBS) separately. This is then followed by theoretical discussions.

\section{Experimental setup}

Fig. \ref{fig1}(a) depicts the fabrication process of the graphene Hall devices. The graphene/$\text{SiO}_2$/Si samples in the study are purchased from Graphenea. To reduce the time that the graphene was in contact with chemicals during the fabrication, a fabrication process using an image reversal photoresist (TI35ES) was designed. At first, photolithography process was used to obtain the graphene Hall structures using a TI35ES as the positive photoresist. A subsequent lithography is then conducted to produce the electrode pattern for the Hall structures using a TI35ES as the negative photoresist. This is followed by a metal evaporation step to deposit the contacts. As the Hall measurements will be performed in solution, the sample surface except for the graphene Hall structures and the electrical contacts on the right is covered by a lithography process using a bio-compatible epoxy-like photoresist (SU8). To minimize the defects, the samples are annealed at 200 $_{\circ}$C in vacuum. They are then mounted on printed circuit boards with those contacts wire-bonded. Before the measurements, all the contacts except for the Hall structures (including the wire bonds) are covered by epoxy. This technique can be used to fabricate very large Hall structures with a little or no observable defects. Samples with the dimension up to 5.4 mm $\times$ 1 mm can be produced, as the noise is lower for larger liquid gated graphene devices\cite{GFET_noise}. 

The setup of LGHM is illustrated in fig. \ref{fig1}(b), where D and S represent the drain and source, respectively. G is the gate (Ag/AgCl) electrode, and I, $V_G$ and A are constant current source, gate voltage source, and ammeter, respectively. The gate voltage source and ammeter are provided by a Keithley 6487 Picoammeter/Voltage Source, while the constant current source is a Keithley 6220 Current Source. Fig. \ref{fig1}(f) shows a photo of the setup used in our experiments. The graphene Hall device and solution are contained in a tube with a cap. Changes in molecular concentration due to evaporation were discounted because the measurement time was very short.

Fig. \ref{fig1}(c) shows a schematic of the graphene Hall structure. $V_{xx}$ is measured between contacts 3 and 4 while $V_{xy}$ between 1 and 3. $I_{DS}$ is the current between the drain and the source contacts, and it is set by the constant current source. Therefore, ${\rho_{xx}} = \frac{{{V_{xx}}}}{{{I_{DS}}}} \cdot \frac{W}{L}$ is the sheet resistance of graphene, where $W$ and $L$ are the width and the length of the Hall bar. In this study, $W=L=200\mu m$. ${R_H} = \frac{{{V_{xy}}}}{{{I_{DS}}B}}$ is the Hall resistance, where $B$ is the magnetic flux density.

As shown in fig.\ref{fig1}(b), $I_{GS}$, as the current between the gate and the source electrodes, is measured by the ammeter, while the voltage $V_G$ is being applied. This is a typical two-electrode electrochemical system, where $V_G$ has been swept back and forth from -0.6 V to 0.6 V at the rate of 4 mV/s, as plotted in fig.\ref{fig1}(d). Note there is a change in the sweeping rate from around 400 s to 490 s, due to the auto-change of the range of the power source. This incurs an abrupt change in the $I_{GS}$-$V_G$ curve (cyclic voltammogram), as plotted in fig. \ref{fig1}(e) where the red curve is the raw data and the blue one is smoothened by performing fast Fourier transform. This, however, will not affect the Hall measurement.

\section{Results}

\begin{figure}
	\includegraphics{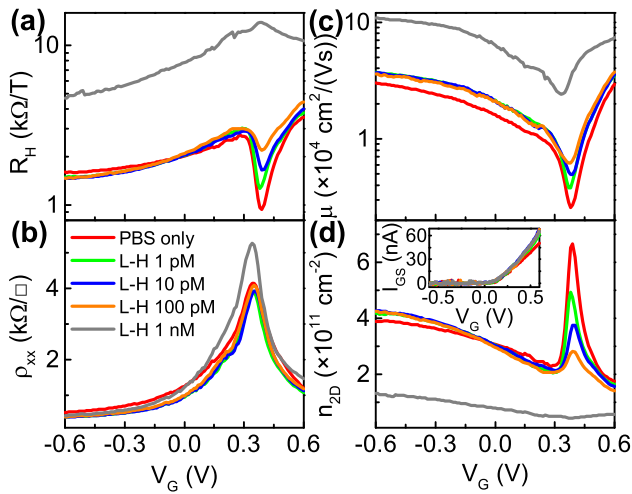}
	\caption{Detection of L-Histidine (L-H) by liquid gated Hall measurement of graphene at pM concentration with $PBS\:(10^{-2}X)$ as the supporting electrolyte. \textbf{(a)}, Hall resistance $R_H$, \textbf{(b)}, sheet resistance $\rho_{xx}$, \textbf{(c)}, mobility $\mu$, and \textbf{(d)}, total carrier density $n_{2D}$ as a function of gate voltage $V_G$ (Inset: gate-source current $I_{GS}$ as a function of $V_G$).
		\label{fig2}}
\end{figure}

Fig. \ref{fig2} shows the measured electronic properties of graphene in solutions containing l-histidine at different concentrations in the pM range by LGHM, where PBS ($10^{-2}X$) is used as the supporting electrolyte. PBS at the concentration of 1X has the pH value of 7.4, and it contains NaCl of 137 mM, KCl of 2.7 mM, $Na_2HPO_4$ of 10 mM, and $KH_2PO_4$ of 1.8 mM. Note that most of the changes in fig. \ref{fig2}(a), (c) and (d) occur near a position where $V_G=V_D \approx 0.35 V$. This is approximately the ``Dirac'' point represented by the peak in fig. \ref{fig2}(b), showing the transition between the hole-dominant region ($V_G<V_D$) and the electron-dominant region ($V_G>V_D$) in graphene. 

As shown in fig. \ref{fig2}(a), $R_H$ gradually increases with the concentration of l-histidine at the gate voltage ($V_G \approx 0.39 V$) slightly above the ``Dirac'' point. This indicates that the measurement of $R_H$ can be used as an effective method to detect l-histidine of these concentrations.

Note the sheet resistance ($\rho_{xx}$) remains almost unchanged for histidine of all concentrations in pM range, as shown in fig. \ref{fig2} (b). Since the transistor-based bio-sensors detect molecules by the change of resistance, this indicates that the sensitivity is not as good as LGHM. In other words, the additional information provided by LGHM is able to improve the sensitivity of the devices based on simple resistance measurement. It should also be noted that there is no evident peak in the $I_{GS}$-$V_G$ curve, as shown in the inset of fig. \ref{fig2}(d). This suggests that the detection of pM L-histidine in PBS by CV using an unfunctionalized graphene electrode is not possible. LGHM therefore provides better sensitivity than the conventional amperometric and potentiometrc methods at low concentration of l-histidine.


The mobility and carrier density shown in fig. \ref{fig2}(c) and (d) are calculated from $\rho_{xx}$ and $R_H$ by the conventional semiconductor model\cite{solid}, where $\rho_{xx} = \frac{1}{en\mu}$ and $R_H = \frac{1}{en_{2D}}$. It is interesting to note that, when $V_G \approx 0.35 V$, the mobility increases with the molecular concentration, while the carrier density decreases.

\begin{figure}
	\includegraphics{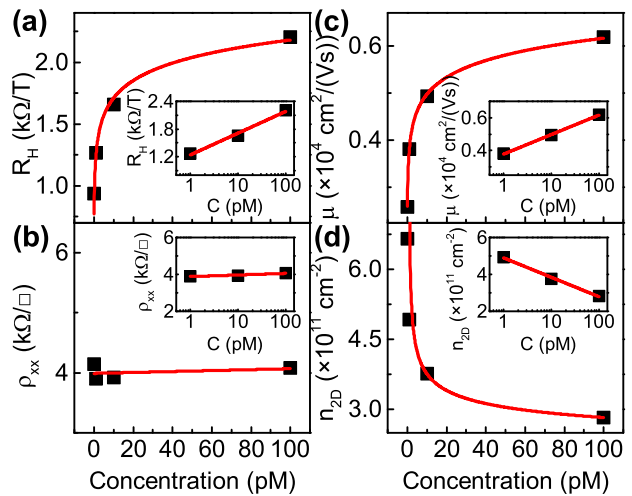}
	\caption{\textbf{(a)}, Hall resistance $R_H$, \textbf{(b)}, sheet resistance $\rho_{xx}$, \textbf{(c)}, mobility $\mu$, and \textbf{(d)}, total carrier density $n_{2D}$ as a function of L-histidine concentration ($C$) in the pM range when the gate voltage is close to the ``Dirac'' point, respectively (Insets: plots in logarithm scale). Black squares: experimental data points. Red lines: fitting curves (logarithm fit for (a), (c) and (d), and linear fit for (b) and the insets).
		\label{fig3}}
\end{figure}

Fig. \ref{fig3} summarizes the changes in the electronic properties of graphene in pM concentration analytes near the ``Dirac'' point. $R_H$, $\mu$, and $n_{2D}$, vary with the concentration in logarithm rules, while $\rho_{xx}$ remains constant. The logarithm relation between electronic properties of graphene with analyte concentration is also observed in other histidine detection experiments\cite{sensing1,sensing3}. It can be seen from fig. \ref{fig3}(c-d) that the mobility/density rises/decays exponentially with the concentration, this keeps $\rho_{xx}$ constant.

There are two more interesting phenomena which should be pointed out. First of all, as mentioned earlier, most of the changes of interest occur at the ``Dirac'' point where electrons and holes co-exist. Secondly, it is worth note that the shape of $R_H$ changes significantly with the analyte concentration, as can be seen in fig. \ref{fig2}(a). This indicates that the symmetry between the properties of electrons and holes might be broken, as the analyte concentration increases. For example, in 1 nM l-histidine solution, the valley in $R_H$ which was previously observed at a position near the ``Dirac'' point in the pM range completely disappeared. This is because l-histidine is a positively charged molecule with the electric dipole moment of 3.6 D. It is attracted to the Hall structure when graphene is negatively biased, i.e. $V_G>V_D$. Since ions in close proximity to graphene introduce charged impurities and then affect the electronic properties\cite{PNAS_Adam}, hence $\mu$ and $n_{2D}$ shown in fig. \ref{fig3}(c-d) vary with the concentration of l-histidine at a positive gate voltage. This sensing mechanism will be discussed in detail later. 

\begin{figure}
	\includegraphics{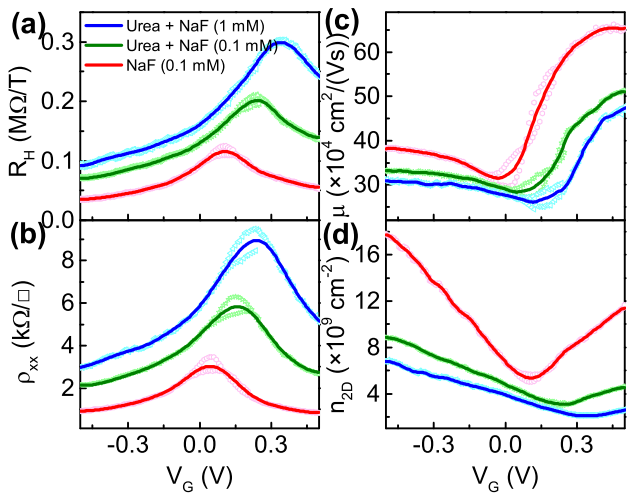}
	\caption{Detection of urea by liquid gated Hall measurement of graphene at high concentration with $NaF\:(100\:\mu M)$ as the supporting electrolyte. \textbf{(a)}, Hall resistance $R_H$, \textbf{(b)}, sheet resistance $\rho_{xx}$, \textbf{(c)}, mobility $\mu$, and \textbf{(d)}, total carrier density $n_{2D}$ as a function of gate voltage $V_G$. Note the empty circles and the solid lines are the raw and the smoothed data (by fast Fourier transform), respectively.
		\label{fig4}}
\end{figure}

It should also be noted that when the concentration reaches 1 nM, changes in both the Hall and the sheet resistances are observed. This indicates that both the transistor-based biosensor and LGHM are sensitive in high concentration electrolyte. 

To further confirm this observation at higher concentration, urea solutions in the $\mu\text{M/mM}$ range are used, where NaF ($100 \mu \text{M}$) is the supporting electrolyte. As can be seen from fig. \ref{fig4}(b), an evident change in $\rho_{xx}$ is observed with the concentration, which indicates the sensitivity of transistor-based bio-sensors in high concentration. As detecting high concentration urea is also possible by CV, the sensitivity of LGHM is therefore believed comparable to the conventional bio-sensing techniques in this case. Note that the shape of $R_H$ is greatly different from that in fig. \ref{fig2}(a), similar to previous observation. This indicates that more study on $R_H$ is necessary.

Moreover, the ``Dirac'' point represented by the peak of $\rho_{xx}$ in fig. \ref{fig4}(b) moves to the direction where $V_G$ becomes more positive. This indicates that either the graphene is being more p-type doped (because it needs more electrons to bring the Fermi level to the ``Dirac'' point) or the influence of the charged impurity on graphene sample becomes greater\cite{PNAS_Adam}. 

However, since there are two amino groups ($NH_2$) in an electrically neutral urea molecule ($CO(NH_2)_2$), the graphene would be n-type doped\cite{RCS_Sun}. Therefore, the phenomenon that the ``Dirac'' point moves to the right direction indicates that the charged impurity scattering plays a more important role. As shown in fig. \ref{fig4}(d), when the urea concentration increases, the minimum value plateau of the charge carrier density curve at ``Dirac'' point gets smaller and wider, and the gradient of the curve decreases. These phenomena have been predicted by a transport theory that included the charged impurity scattering in graphene\cite{PNAS_Adam}.

Due the presence of the low carrier density, which is determined by the screened, charged impurity potential\cite{PNAS_Adam}, the mobility increases if the sheet resistance does not change too much. This also means a very high mobility can be obtained by engineering the electrolyte concentration, which is interesting for nano-electronics. It can be seen in fig. \ref{fig4}(c) the highest mobility in our experiments reaches a value at around $6 \times 10^5 cm^2/(Vs)$, which is higher than most of the reported values.

%
\section{Discussion of the sensing mechanism}

\begin{figure}
	\includegraphics{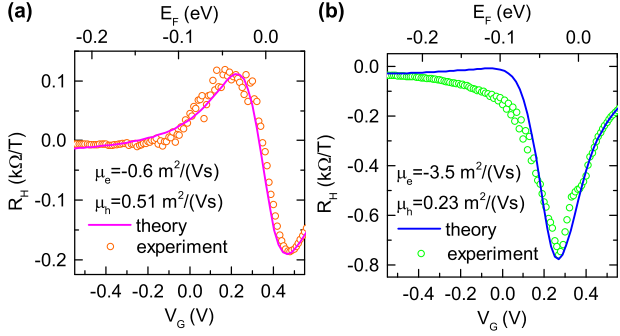}
	\caption{Measured Hall resistance ($R_H$) of graphene in \textbf{(a)}, $H_2O$ and \textbf{(b)}, $PBS\:(10^{-3}X)$. Empty circles: experimental results, solid lines: theoretical fittings. Note that the Fermi level ($E_F$) is used in the calculations, instead of gate voltage.
		\label{fig5}}
\end{figure}

As discussed earlier, the change of $R_H$ near the ``Dirac'' point may reflect the break of symmetry of the properties between electrons and holes, and it could account for the sensing mechanism. $R_H$ of graphene is then measured in DI water and in a solution containing PBS (10$^{-3}$ X) separately, as indicated by the empty circles in fig. \ref{fig5}.

It can be seen that $R_H$ has a sinusoidal-like shape in DI water. This is consistent with the observation in Hall effect measurements for graphene solid-state devices which do not involve liquid\cite{Science_Novo}. A reason for this is that DI water is considered a clean solution with only a few residual ions. Neutrally charged water molecules can be considered as a simple dielectric medium between the gate and graphene, without changing the electron/hole properties differently. When there are more ions in the solution, they act as charged impurities in the environment of graphene. This introduces electron-hole puddles in graphene\cite{PNAS_Adam}, and could affect the electron/hole properties (such as mean free paths, mobility, carrier density, etc.). These properties are changed symmetrically in solid-state devices, as the impurities follow a (symmetric) Gaussian distribution about the ``Dirac'' point\cite{NPhy_Chen}. However, since the ions on the electrode surface are not described by the simple (symmetric) Maxwell-Boltzmann statistics in the presence of electric potential\cite{Zhan_EDL}, the electron/hole properties are affected differently.

The solid curves in fig. \ref{fig5} are the theoretical fittings to $R_H$ using a relation derived from the two-carrier model of graphene\cite{PRB76_Hwang}
\begin{equation} \label{eq1}
	R_H = \frac{n_h\mu_h^2[1+(\mu_eB)^2]-n_e\mu_e^2[1+(\mu_hB)^2]}{e[(\mu_e\mu_hB)^2(n_e-n_h)^2 + (n_e\mu_e-n_h\mu_h)^2]},
\end{equation}
where the subscripts $h$ and $e$ denote holes and electrons, respectively, and $B$ is the magnetic flux density. Note that electron/hole mobility can have different values in this equation. The fittings to $R_H$ in fig. \ref{fig5} show that the electron-hole mobility becomes more asymmetric in PBS ($\mu_e = -3.5 m^2/(V\cdot s)$ and $\mu_h = 0.25 m^2/(V\cdot s)$) than that in DI water ($\mu_e = -0.6 m^2/(V\cdot s)$ and $\mu_h = 0.51 m^2/(V\cdot s)$). This confirms the previous inference where the symmetry between electron-hole properties should be broken when the analyte concentration increases.

This asymmetry comes from the interaction between the solvents and graphene. Depending on the charge of the ions, dipole moment of the molecules, and how their energy levels (such as HOMO and LUMO) interact with graphene, the electronic properties will be changed differently by charged impurity scattering, doping process, and so on. Since the induced charged impurity scattering event is not necessarily the same for electrons and holes, the symmetry of $R_H$ curve will change for different electrolyte of various concentration. This phenomenon has been consistently observed in the experiments and it is used as the mechanism to detect the change of the charged impurity scattering induced by solvated ions in solution.

\begin{figure}
	\includegraphics{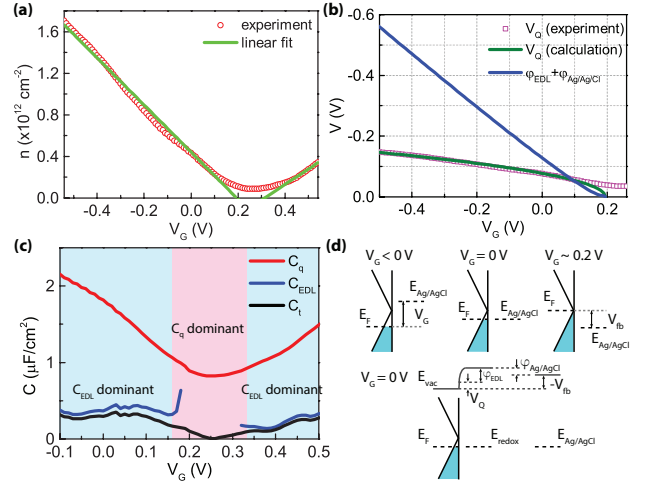}
	\caption{Determination of the potential profile and capacitance values in the graphene-electrolyte system (PBS, 10$^{-3}$ X). \textbf{(a)}, Carrier density as a function of the gate voltage (empty circles: experimental data, solid lines: fitted lines). \textbf{(b)}, Potential drops within graphene ($V_Q$) and electrolyte ($\varphi_{EDL} + \varphi_{Ag/AgCl}$) at different gate voltages. \textbf{(c)}, Calculated quantum, EDL, and total capacitance values. \textbf{(d)}, Band diagrams of the graphene-electrolyte system at different gate voltages. The bottom one describe all the quantities ($V_Q$, $\varphi_{EDL}$, and $\varphi_{Ag/AgCl}$) in detail.
		\label{fig6}}
\end{figure}

Further analysis of the system can be performed using the data obtained in fig. \ref{fig5}(b), in particular, the quantum capacitance and the electrical double layer capacitance. This is because different conclusions on the importance of the quantum capacitance (or the potential fill) of graphene when it is in contact with solution are reported for various systems\cite{QC_G,QC_FET}. Studying the quantum capacitance and also the energy level diagram is helpful for better understanding the system in future.

A rough estimation of the charge carrier density using the aforementioned conventional semiconductor model is described by the empty red circles in fig. \ref{fig6}(a). Note that the minimum plateau of the carrier density at around $V_G \approx 0.27 V$ do not represent the actual value of $n$, as this model is invalid for graphene when close to ``Dirac'' point. The green lines are the linear fits to the data away from the minimum plateau. They are expected to cross at zero carrier density ($n=0$) at ``Dirac'' point, as there are no carriers at the ``Dirac'' point. This `inconsistency' could be attributed to the asymmetry of the electron-hole mobilities and the formation of electron-hole puddles.

The potential drops in graphene ($V_Q$) are derived from the definition of carrier density, where $n \approx  \frac{e^2}{\pi (\hbar v_F)^2} V_Q^2$. The purple squares and the olive curve in fig. \ref{fig6}(b) are calculated using the empty circles and green lines in (a), respectively. The electrical double layer potential ($\varphi_{EDL}$) is then obtained by a relation $V_G=V_{fb}+V_Q+(\varphi_{EDL}+\varphi_{Ag/AgCl})$, where $\varphi_{Ag/AgCl}$ is a constant voltage difference between the electrolyte and the Ag/AgCl gate electrode, and $V_{fb}$ is the ``flat-band'' voltage indicated in fig. \ref{fig6}(d). It can be seen that $V_Q$ becomes 0 at a point close to $V_G\sim 0.2V$, which is the ``Dirac'' point given by the linear fit. This is because a zero carrier density at the ``Dirac'' point suggests a zero $V_Q$ according to the definition, as represented by the olive curve in fig. \ref{fig6}(b). 

It is also worth noting that the slope of the blue curve is steeper than that of the olive curve when $V_G$ is far away from the ``Dirac'' point, for example when $V_G<0.1V$. However, this value becomes smaller than the slope of the olive curve when $V_G$ is close to the ``Dirac'' point. This information is particularly helpful for estimating the capacitance values, as the quantum ($C_q$), electrical double layer ($C_{EDL}$), and total ($C_t$) capacitances are defined as $C_q = e\frac{\mathrm{d}n}{\mathrm{d}V_Q}$, $C_{EDL} = e\frac{\mathrm{d}n}{\mathrm{d}\varphi_{EDL}}$, and $C_t = e\frac{\mathrm{d}n}{\mathrm{d}V_G}$.

The calculated capacitance values are shown in fig. \ref{fig6}(c). Note that $C_{EDL}$ is derived using the blue curve in fig. \ref{fig6}(b), and this accounts for the discontinuity in $C_{EDL}$ in fig. \ref{fig6}(c). In spite of this, we are still able to estimate the role of different capacitances. Since both $C_q$ and $C_t$ show a minimum at a point around $V_G\approx 0.25V$, and $C_{EDL}$ intend to increase in this region, $C_q$ is believed dominating the total capacitance in the red regime in fig. \ref{fig6}(c). In the blue regime where $V_G$ is far away from 0.25 V, $C_t$ changes with $C_{EDL}$. This means that $C_t$ is dominated by $C_{EDL}$.

A detailed band diagram of the system at different gate voltage is shown in fig. \ref{fig6}(d). As $C_q$ is only dominating close to the ``Dirac'' point and hence, the energy level in graphene changes more than that in EDL. Therefore, the electronic properties in graphene respond more quickly to the changes in the environment in this range. This is also associated with the observable changes in $R_H$ and $n_{2D}$ close to the ``Dirac'' point in fig. \ref{fig2}.

\section{Conclusion}

Liquid gated Hall effect measurement is introduced in this work to detect molecules in low concentration electrolyte. The obtained $R_H$ shows an ultra-sensitivity for l-histidine in pM concentration, while there are no observable responses by amperometric and potentiometric methods. It indicates that this extra information provided by LGHM is able to improve the sensitivity of conventional methods in some cases. The derived properties such as $n_{2D}$ and $\mu$ also show clear changes in response to the change of the solution concentration. Therefore, there is a great potential of the method that we can choose different properties to get the best sensor response, which makes this technique more sensitive in some cases than simple transistor based sensors. Our study shows that the asymmetric electron-hole mobility induced by the charged impurities in solution serves as the sensing mechanism for LGHM. Moreover, the quantum capacitance of graphene is found only dominant near the ``Dirac'' point in out system.

Another advantage of LGHM is that solvated ions within the solution are forcefully attracted towards the sensors due to the presence of the electric field. Therefore it might be possible to achieve fast detection using this method. While in conventional potentiometric sensors, molecules are mostly driven by diffusion which is a much slower process\cite{NL_dl,PRL_dl,NB_dl}. Although electric field is also present in the solution for amperometric methods, the sensitivity at low concentration is reduced as reactants can be rapidly consumed at the electrode beyond the diffusion rate, in the extreme case, this will lead to a reverse reaction\cite{echem}. 

However, there are several limitations of this method which should be considered in future development. As shown in fig. \ref{fig1}(b) and (f), the setup of this method is more complicated than either CV or ISFET. However, it is believed a smaller device can be developed due to the presence of portable Hall measurement equipment. Similar to CV, another disadvantage is that some molecules may stay on the surface of graphene after detection. This could be resolved by some techniques such as graphene annealing, CV cleaning, and so on. The third one is the interaction between analyte molecules and graphene requires more study, such as first-principal calculation. This might not be very critical for application, if one decides to functionalize graphene. In spite of these limitations, the strikingly good sensitivity provides a promising future for utilizing this technique in application of molecular sensing.









%

\end{document}